\documentclass[letterpaper, 10 pt, conference]{ieeeconf}  

\IEEEoverridecommandlockouts
\usepackage{amssymb,latexsym,amsmath}     
\usepackage{graphics} 
\usepackage{subcaption}
\usepackage{graphicx}
\usepackage{epsfig} 
\usepackage{float}
\usepackage{amsmath}
\usepackage{arydshln}
\usepackage{float}
\newenvironment{definition}[1][Definition]{\begin{trivlist}
\item[\hskip \labelsep {\bfseries #1}]}{\end{trivlist}}

\title{\LARGE \bf
Modelling and controllability of the motion of a slender, flexible micro-swimmer}

\author{Sudin Kadam$^{1}$ and Ravi N. Banavar$^{1}$
\thanks{$^{1}$Sudin Kadam and Ravi N. Banavar are with Systems and Control Engineering Department, Indian Institute of Technology Bombay, Mumbai, India, 400072
        {\tt\small sudin@sc.iitb.ac.in, banavar@iitb.ac.in}}}

\begin{document}
\maketitle
\thispagestyle{empty}
\pagestyle{empty}

\begin{abstract}
The mechanism of swimming at very low Reynolds number conditions is a topic of interest to biologists and engineering community. We develop a novel kinematic model of a slender flexible swimmer which locomotes in a low Reynolds number regime. In contrast to existing techniques that model such systems as a connected set of straight, rigid links, the novelty of our technique stems from the fact that we model the swimmer with two components - one is a straight, rigid body (the head) and the other is a flexible member (the tail). Using Cox theory we model the gradient of the forces as a function of the instantaneous shape of the swimmer and its velocity. By virtue of the low inertia conditions, an expression for the translational and rotational velocity of the head is obtained for the planar motion in the form of a Lie algebra of the Special Euclidean group. We explain the principal fiber bundle structure of the configuration space of the swimmer and use that to show a weak controllability result for a type of slender flexible swimmer where the shape space is the space of all continuous curves of a given length. A set of simulation results is presented showing the variation of the swimmer head velocity for a bump function moving along the swimmer length.
\end{abstract}
\section{INTRODUCTION}
\begin{figure}[!htb]
    \centering
    \includegraphics[width=0.4\textwidth]{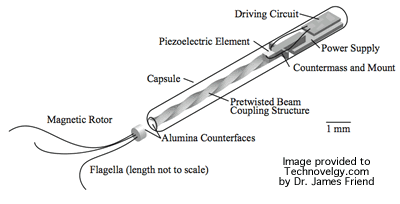}
    \caption{A flagellum propelled microbot \cite{Technovelgy}}\label{fig:micromotor}
\end{figure}

Swimming at micro scales is a topic of growing interest. There has been a lot of research in exploring new and efficient ways to generate propulsion at these scales, see \cite{becker2003self}, \cite{dreyfus2005microscopic}. A better understanding of the mechanism of swimming can lead to many applications in  several fields such as micro-machining, nano technology and medicine for targeted drug delivery. Micro-robotics is one of the recently evolving fields in miniature robotics, especially mobile robots with characteristic dimensions to the scale of a micron. Figure \ref{fig:micromotor} shows an example of a microswimmer developed at Monash University. Apart from applications in robotics, since cell or microbial locomotion is an essential part of biological systems, understanding the means of locomotion in this regime is of a great academic interest to biologists.

Most of our intuition about locomotion stems from inertia-dominant systems used by larger animals. As opposed to such systems, microorganisms resort to a creeping motion. This motion occurs in a fluid medium with very low Reynolds number, which is the ratio of the inertial to viscous forces acting on the swimmer's body. The essential mechanism of the motion in this regime is based on the fact that the inertial effects are almost negligible as compared to the viscous forces due to Reynolds number of the order of $10^{-4}$. To get a relative sense of the numbers, the Reynolds number for a man swimming in water is of the order of $10^4$ \cite{najafi2004simple}, \cite{cohen2010swimming}. Organisms such as motile bacteria and  Escherichia Coli, commonly found in the human intestines, use one or more flagella, which are long helical filaments and act as propeller. Figure $1$ shows a few such microbes with flagella.
\begin{figure}[!htb]
\begin{subfigure}{0.23\textwidth}
  \centering
  \includegraphics[width=0.65\linewidth]{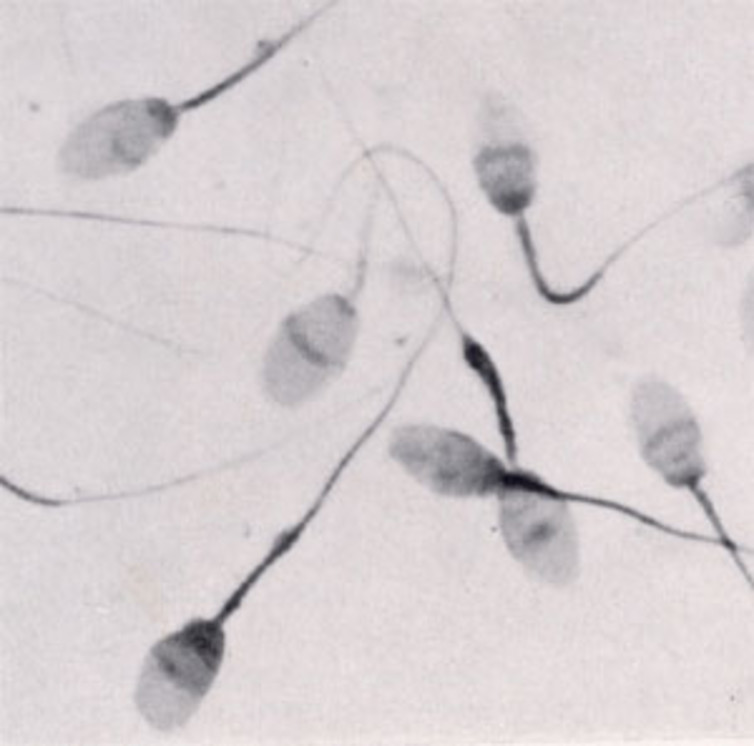}
  \caption{Sperm cells \cite{sperm}}\label{fig:sperm1}
  \end{subfigure}
\begin{subfigure}{0.23\textwidth}
\centering
  \includegraphics[width=0.95\linewidth]{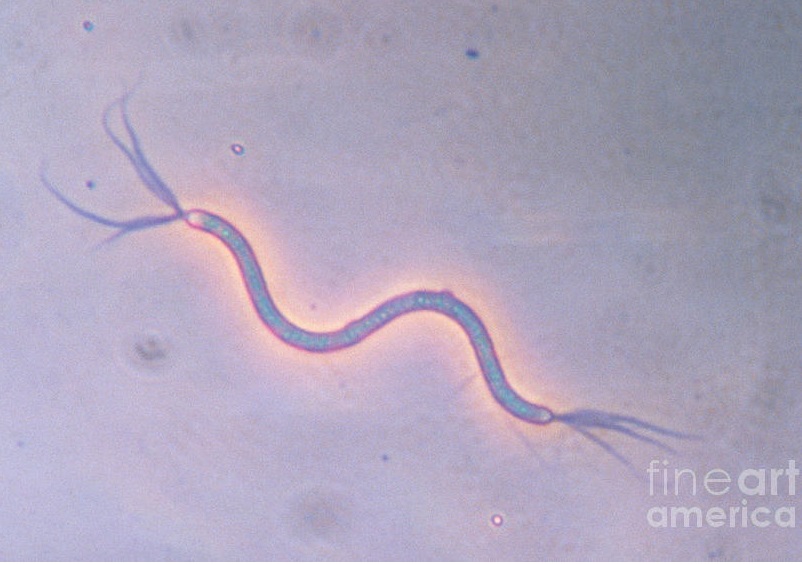}
  \caption{Spirillum bacteria \cite{spirrilum}}\label{fig:spirillum_volutans }
\end{subfigure}
\end{figure}

There has been a lot of research on the mechanism of motion of micro-organisms and micro-robotic swimmers. The Purcell's swimmer is a famous example of the simplest possible low Reynolds number swimmer \cite{hatton2013geometric}, \cite{purcell1977life}. This swimmer has three straight, slender and rigid links connected through 2 rotary joints through which the actuation is affected. Through non-reciprocal motion of the outer links, the Purcell's swimmer can achieve any point-to-point motion on its configuration space. Although there has been a lot of research on the Purcell's swimmer and a few other low Reynolds number swimmers which have finitely many control inputs from mechanics and control theoretic perspective \cite{cohen2010swimming}, \cite{hatton2013geometric}, \cite{berg2008coli}, \cite{kadam2016geometric}, the dynamic behaviour of these swimmers differ significantly from most of the micro-organisms found in the nature whose limbs are not rigid links. Most of the micro-organisms have flexible limbs, whose actuation results in the net motion of the swimmers. Clearly, there is a need to understand the mechanism of motion of such microswimmers with flexible link, which is far more challenging than such analysis with rigid links. The approach to model these systems is through Stokes equations.

One of the earlier works formulates this problem of flexible swimmers' modelling in terms of a gauge field on the space of the shapes \cite{shapere1987self}. Existence and uniqueness of the solution to the model obtain using exterior Stokes problem when a deformable body is placed in a three-dimensional fluid is presented in \cite{silvestre2002slow}. The recent works have extensively studied control theoretic problems such as controllability and optimal control of such self propelling deformable bodies \cite{galdi1999steady}, \cite{martin2007control}. All of these works target to solve the problem when the deformable body is a smooth embedding of sphere in the ambient fluid. Our work relates to a similar control theoretic modelling and analysis of flexible micro-swimmers, but which are slender. As discussed earlier, the motivation for this is that a lot of microorganisms found in the nature are slender. The approach makes use of Cox theory for modelling, which makes use of the slenderness of the swimmer to simplify the computation while solving the exterior Stokes problem.

\subsection{Contribution}
To the best of our knowledge, this is the first attempt at getting a mathematical model of a slender, flexible swimmer's motion at low Reynolds number conditions. The model obtained gives an expression for the velocity of the head of the swimmer as a function of the shape and velocity of the rest of the body. For many micro-robotics applications, since the payload of the swimmer is located in the frontal part of the body, having control over the velocity and position of the swimmer's head is of importance. The model presented in this paper is in a form suitable to do an analysis of the swimmer head motion. Furthermore, we present a geometric interpretation of the configuration space of the swimmer. A weak controllability analysis is presented using the strong controllability of the planar Purcell's swimmer which gives a result on the ability to do point to point reconfiguration of the swimmer head. Moreover, a set of simulation results for a class of shape curves and their velocities gives an insight into the relationship of the parameters of these motion curves with the swimmer motion.

\subsection{Organization of the paper}
Section 2 gives a brief explanation of the Stokes equations and Cox theory to get the expression of the force gradient acting on a slender swimmer. This is followed by using this force expression and the condition of strong dominance of the viscous forces over inertial forces to obtain the kinematic model of the swimmer. The geometry of the configuration space of this swimmer along with a result on weak controllability is presented in section 3. Section 4 analyses the behaviour of the kinematic model for a class of spatio-temporal curves of the shape of the swimmer through a set of simulation results.  Section 5 concludes the paper with a discussion on the possible avenues for the future work.

\section{Kinematic model of the flexible swimmer}
\begin{figure}[!htb]
\centering
\includegraphics[scale=.4]{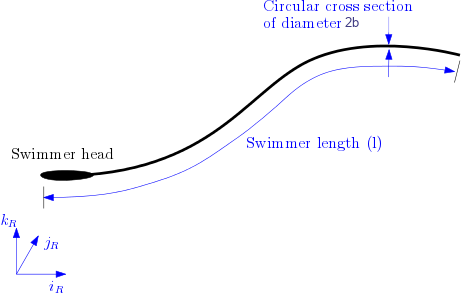}
\caption{Slender flexible swimmer}
\label{FlexibleSwimmerSchematic}
\end{figure}
\begin{figure}[!htb]
\centering
\includegraphics[scale=.6]{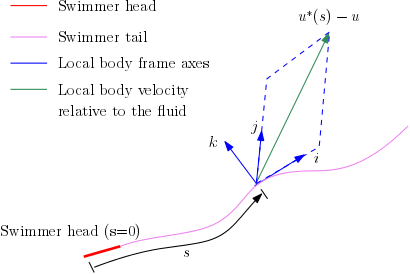}
\caption{Slender flexible swimmer}
\label{FlexibleSwimmerBodyFrame}
\end{figure}

Consider a long slender body of a circular cross-section, the length of the body be $l$ and the cross-sectional radius $b$ along with a reference coordinate frame. Figure \ref{FlexibleSwimmerSchematic} shows an arbitrary configuration of such a slender body in a 3-dimensional space along with the inertial reference frame defined by axes $\{i_R,\:\: j_R, \:\: k_R \}$. We do arc-length parametrization of the body along its length measured from the head. So a point on the body is characterised by the dimensionless parameter $s \in [0,1]$. The cross sectional radius at a given point of the curve is taken as $b \lambda(s)$, where $\lambda(s)$ is a dimensionless quantity function of $s$. Let $u^*(s)$ be the velocity of the point at arc-length $s$ with respect to the reference frame. $u(r)$ be the velocity of fluid at a general point at $r$ with respect to the reference frame. The velocity field $u$ and pressure field $p$ in the fluid satisfy the dimensionless Stokes equations
\begin{equation}
\mu \nabla^2 u - \nabla p = 0, \qquad \nabla \circ u = 0
\end{equation}
with the boundary conditions,
\begin{align*}
u &= u^*(s) \text{ on the body at point s} \text{ and, } \\
u &\rightarrow 0 \text{ as } r \rightarrow \infty
\end{align*}
Where $\nabla$ is the operator defined as $\frac{\partial}{\partial x} i_R +\frac{\partial}{\partial y} j_R+\frac{\partial}{\partial z} k_R$ and $\nabla^2 = \frac{\partial^2}{\partial x^2}+\frac{\partial^2}{\partial y^2}+\frac{\partial^2}{\partial z^2}$ is the Laplacian operator. The complete velocity and pressure field is the solution of the Stokes equations subject to the boundary conditions. To write the expression of the force density at each point of the swimmer we set a reference Cartesian coordinate frame at each point of the swimmer as shown in figure \ref{FlexibleSwimmerBodyFrame}. In these frames, the axis $i$ is along the tangent direction, $j$ is orthogonal to $i$ and in the plane formed by $u^*(s)-u$ and the $i$ axis, and $k$ completes the right handed frame.
\subsection{Cox theory and fluid force gradient}
According to Cox theory \cite{cox1970motion}, the density of the fluid forces acting at a point at arc-length $s$ are obtained in this local coordinate frame in terms of the swimmer's local shape, its slenderness ratio $h = \frac{b}{l}$ and the difference between the local body velocity and fluid velocity in the following form.
\begin{equation}\label{ForceDensity1}
F(s) = 2 \pi ( E(h,s) i - 4 C(h,s) j -4 D(h,s) k )
\end{equation}
where,
\begin{align*}
C(h,s) &= \frac{1}{(log (h))^{-1}} (u_j - u_j^*(s)) + \frac{1}{(log (h))^{-2}} \{ \frac{1}{2} (u_j - \\ \nonumber
& u_j^*(s)) (\frac{1}{2} + log (2) - log(\lambda) + log(\epsilon)) \} + \frac{1}{2} J_j + ... \\
D(h,s) &= \frac{1}{2 (log (h))^{-2}} J_k \\
E(h,s) &= -\frac{1}{(log(h))^{-1}} (u_i - u_i^*(s)) + \frac{1}{(log(h))^{-2}} \{ \frac{1}{2} (u_i - \\ \nonumber
& u_i^*(s)) (\frac{1}{2} + log(2) - log(\lambda) + log(\epsilon)) \} + \frac{1}{2} J_i + ...
\end{align*}
and $J$ is a vector given by 
\begin{align}
J_i &= \frac{1}{2} [ \int^{s-\epsilon}_0 + \int^1_{s+\epsilon}] \{ \frac{\delta_{ij}}{|R - \hat{R}|} +  \frac{(R_i - \hat{R}_i)(R_j - \hat{R}_j)}{| R - \hat{R}  |^3}  \times \nonumber \\ 
& \qquad \qquad \{ \delta_{jk} - \frac{1}{2} \hat{t}_j \hat{t}_k  \} \{u_k(\hat{R}) - u^*_k(\hat{s})  \} \} d\hat{s}
\end{align}
where $\delta_{ij}$ is the Kronecker delta defined as 
\begin{equation}
\delta _{{ij}}={\begin{cases}0&{\text{if }}i\neq j,\\1&{\text{if }}i=j.\end{cases}}
\end{equation}
and $R$ is the position vector of the center-line of the infinitesimal segment of length $2\epsilon$ and $\hat{R}$ is the position $r$ at the point on the center line with $s = \hat{s}$, both in the inertial frame. $\hat{t}$ is the unit tangent vector to the curve formed by the swimmer at arc length $\hat{s}$ in the inertial frame. The force gradient equation (\ref{ForceDensity1}) can be compactly written as the following vector equation in terms of the tangent vector $\hat{t}$ at the arclength $s$  to the curve formed by the swimmer's body \cite{cox1970motion}
\begin{align}\label{Force_eps}
&F(s) = 2\pi [ [\frac{u - u^*(s)}{log(k)} + \frac{J + (u-u^*(s)) log (2) \frac{\epsilon}{\lambda})}{(log(h))^2}]\circ \nonumber \\ 
& \:\:\: [\hat{t} \hat{t} -2 \mathbb{I}] + \frac{(u-u^*)}{(log(h))^2}\circ[3 \hat{t}\hat{t} -2\mathbb{I}] + O\{ \frac{1}{(log(h))^3} \} ]
\end{align}
where, $\mathbb{I}$ is the identity operator defined such that for any vector in $v \in \mathbb{R}^3$, $v \circ \mathbb{I} = v$. For $\epsilon \rightarrow 0$, the value of $J$ can be written as 
\begin{equation}
J = - \{ u(R) -u^*(s) \} log(\epsilon) + O(1)
\end{equation}
We substitute this value of $J$ in the equation \ref{Force_eps}. We also assume that the ambient fluid is stationary, $u = 0$, and that body has a circular cross-section of a constant radius, i.e. $\lambda(s) = 1$. A a consequence, we get the following form of force gradient which is independent of $\epsilon$
\begin{align}\label{Force_density}
& F(s) = 2\pi [ [\frac{- u^*(s)}{log(h)} + \frac{-u^*(s) log(2) }{(log(h))^2}]\circ[\hat{t} \hat{t} -2 \mathbb{I}] - \nonumber \\
& \qquad \qquad \frac{-u^*(s)}{2(log(h))^2} \circ [3 \hat{t}\hat{t} -2I] + O\{ \frac{1}{(log(h))^3} \} ]
\end{align}

\subsection{Model of the swimmer with head and tail}
We now obtain the expression for the total force and moment acting on the swimmer body with respect to the reference frame at the head of the swimmer $(s=0)$. We refer to this frame as the head frame in the rest of the paper. We define $Q(s)$ to be an appropriate transformation which transforms the forces and moments in the body frame at the point at arc-length $s$ to those in the head frame, and $r(s)$ is the position vector of the point at arc-length $s$ with respect to the head frame. Thus the net force and moment in the head frame is obtained by integrating the force gradient from the equation \ref{Force_density} over the entire body length as follows -
\begin{align}
F &= 2 \pi \int_0^1 Q(s)F(s) ds \\
M &= 2 \pi \int_0^1 r(s) \times [Q(s)F(s)] ds
\end{align}
By virtue of the low Reynolds number conditions, the net forces and moments acting on the body should be zero in the body coordinate frame at the tip of the head $(s=0)$. Along with this, we substitute $\frac{1}{log (h)} = c$ and neglect higher order terms in equation \ref{Force_density} to get the following net force and moment expressions in the head frame -
\begin{align}
& 2 \pi\int_0^1 Q(s) \left[ -c  u^*(s) -  log (2) c^2 u^*(s) \circ [\hat{t} \hat{t} -2 \mathbb{I}] - \right. \nonumber \\
&\left. \qquad  u^*(s) c^2 \: \circ  [3 \hat{t}(s)\hat{t}(s) -2I] \right] ds = 0 \\
& 2 \pi \int_0^1 Q(s) \left[ r(s) \times [-c  u^*(s) -  log (2) c^2 u^*(s)\circ[\hat{t}(s) \hat{t}(s) - \right. \nonumber \\
&\left. \qquad 2 \mathbb{I}] - u^*(s) c^2 \circ [3 \hat{t}(s) \hat{t}(s) -2\mathbb{I}]] \right] ds = 0
\end{align}
We now assume that the swimmer's head is a small rigid straight link of length $\delta$ and of the same cross sectional radius as that of the rest of the body. We split the forces and moments acting on the body as those acting on the head and the tail to get the force and moment balance equations as follows -
\begin{align}
& F_{head} + F_{tail} = 0 \\
& M_{head} + M_{tail} = 0 
\end{align}
\begin{figure}
  \begin{center}
\includegraphics[scale=0.45]{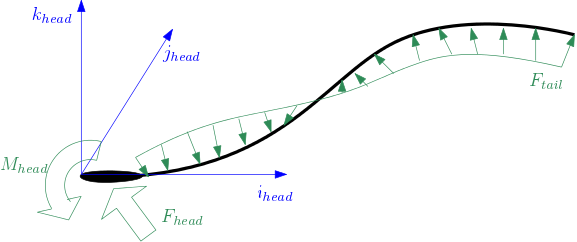}
\caption{Forces and moments on the swimmer}
\label{swimmer_forces_moments}
  \end{center}
\end{figure}
The forces and moments are represented in the frame attached to the tip of the swimmer head. Figure \ref{swimmer_forces_moments} shows the schematic. For the further analysis we consider the motion of the swimmer restricted in the plane formed by the $x, \: y$ axes of the head frame. We consider the tip of the head to have translational velocity as $v = [v_{0x}, \: v_{0y}, \: 0]^T$ with respect to the reference frame and angular speed of this segment as $\omega_0$ about this tip with respect to the reference frame at the head. We note that the transformation $Q(0)$ for the head segment is just the identity transformation. Thus we get the following force equation for the motion of the head of the swimmer -
\begin{align}
& \int_0^{\delta} \left[ -c  (v+ s\hat{t}(s) \times \omega) -  log (2) c^2 (v+ s\hat{t}(s) \times \omega)\circ \right. \nonumber \\
&\left. \:\:\:\: [\hat{t}(s) \hat{t}(s) -2 \mathbb{I}] - c^2 (v+ s\hat{t}(s) \times \omega)\circ[3 \hat{t}(s)\hat{t}(s) -2\mathbb{I}] \right] ds \nonumber \\
& = -\int_{\delta}^1 Q(s) \left[ -c  u^*(s) -  log (2) c^2 u^*(s) \circ [\hat{t}(s) \hat{t}(s) -2 \mathbb{I}] - \right. \nonumber \\
&\left. \qquad c^2 u^*(s)\circ[3 \hat{t}(s)\hat{t}(s) -2\mathbb{I}] \right] ds \label{Force1}
\end{align} and the moment equation is obtained as follows-
\begin{align}
& \int_0^{\delta} r(s) \times [-c  (v+ s\hat{t} \times \omega) -  log (2) c^2 (v+ s\hat{t} \times \omega) \circ  \nonumber \\
& \qquad [\hat{t}(s) \hat{t}(s)-2 \mathbb{I}] - c^2 (v+ s\hat{t} \times \omega)\circ[3 \hat{t}(s)\hat{t}(s) -2\mathbb{I}]] ds \nonumber \\
&= -\int_{\delta}^1 \left[ r(s) \times Q(s) [-c  u^*(s) -  log (2) c^2 u^*(s) \circ  \nonumber \right. \nonumber \\
&\left. \qquad [\hat{t}(s) \hat{t}(s) -2 \mathbb{I}] - c^2 u^*(s) \circ [3 \hat{t}(s)\hat{t}(s) -2\mathbb{I}]] \right] ds \label{Moment1}
\end{align}
The objective is to first determine the translational and rotational velocities of the swimmer head $v$ and $\omega$ for a given profile of $u^*(s) , \: s\in (\delta,1]$, and then to design $u^*(s,t)$ to achieve  a desired velocity trajectory $v(t)$ of the head of the swimmer. Using these equations we will now obtain the head velocity of the swimmer as a function of the shape position and shape velocity $u^*(s)$. We consider the planar motion. Thus at each point the z component of the forces acting on the body is zero and only the moment about the z-axis is non-zero. We denote by $v_0=[v_{0,x}, \:v_{0,y}]^T$ the translational velocity of the tip of the swimmer head $(s=0)$ and $\omega_0$ as the angular velocity of the swimmer head about the its tip. Let $\hat{t}_0$ is the tangent vector to the head. Thus the left hand side of the force equation \ref{Force1} is computed as follows
\begin{align}
& \int_0^{\delta} \left[ -c (v_0 + s \hat{t}_0 \times \omega_0) - log(2)c^2 (v_0 + s \hat{t}_0 \times \omega_0) \circ \right. \nonumber \\
&\left (\hat{t}_0 \hat{t}_0 - 2 \mathbb{I}) - (v_0 + s \hat{t}_0 \times \omega) \circ (3 \hat{t}_0 \hat{t}_0- 2 \mathbb{I}) \right] ds \nonumber \\
&=-c \delta \begin{bmatrix}
v_{0x} \\
v_{0y}
\end{bmatrix} + \omega_0 \frac{\delta^2}{2} \begin{bmatrix}
t_{0y} \\
-t_{0x}
\end{bmatrix} - log(2)c^2 \{\delta \begin{bmatrix}
v_{0x} \\
v_{0y}
\end{bmatrix} -  \omega_0 \frac{\delta^2}{2} \times \nonumber \\ 
& \qquad  \begin{bmatrix}
t_{0y}	\\
-t_{0x}
\end{bmatrix}  \} \circ \{ \begin{bmatrix}
t_{0x}	\\
t_{0y}
\end{bmatrix}\begin{bmatrix}
t_{0x}	\\
t_{0y}
\end{bmatrix} - 2 \mathbb{I} \} - c^2 ( \delta \begin{bmatrix}
v_{0x} \\
v_{0y}
\end{bmatrix} +  \nonumber \\ 
& \qquad \frac{\delta^2}{2} \omega_0 \begin{bmatrix}
t_{0y}	\\
-t_{0x}
\end{bmatrix}) \circ \{3 \begin{bmatrix}
t_{0x}	\\
t_{0y}
\end{bmatrix} \begin{bmatrix}
t_{0x}	\\
t_{0y}
\end{bmatrix} -2 \mathbb{I}  \} \nonumber \\
& = \begin{bmatrix}
a_{11}  & a_{12}  &a_{13} \\
a_{21}  & a_{22}  &a_{23}
\end{bmatrix} \begin{bmatrix}
v_{0x} \\
v_{0y} \\
\omega_0
\end{bmatrix}\label{Forces_LHS_final}
\end{align}
where,
\begin{align*}
& t_{0} = [t_{0x}, \: t_{0y}]^T \text{ is the tangent vector at the head } s=0 \\
& a_{11} = - c \delta - c^2 \delta log(2) (t_{0x} t_{0x} + 1) - c^2 \delta (3 t_{0x}t_{0x} - 2) ,\\
& a_{12} =  0, \\
& a_{13} = \frac{\delta^2} {2} (-t_{0y} - c^2 log(2) t_{0y} t_{0x} t_{0x} - c^2log(2) + \\ \nonumber 
& \qquad \qquad \: 3 c^2 t_{0y} t_{0x} t_{0x} - c^2t_{0y} ),\\
& a_{21} = 0, \\
& a_{22} = - c \delta - c^2 \delta log(2) (t_{0y} t_{0y} + 1) - c^2 \delta (3 t_{0y}t_{0y} - 2), \\
& a_{23} =  \frac{\delta^2}{2} ( - t_{0x} - c^2 log(2) t_{0x} t_{0y} t_{0y} + log(2) c^2 t_{0x} - \\ \nonumber 
& \qquad \qquad \: 3 c^2 t_{0x} t_{0y} t_{0y} - 2c^2t_{0x})
\end{align*}
Similarly, the left hand side of the moment equation \ref{Moment1} gets simplified as follows -


\begin{align}
& \int_0^{\delta} s \hat{t}_0 \times [-c (v_0+ s\hat{t}_0 \times \omega_0) -  log(2)c^2 (v_0+ s\hat{t}_0 \times \omega_0) \circ \nonumber \\
& \qquad [\hat{t}_0 \hat{t}_0 - 2 \mathbb{I}] - c^2 (v+ s\hat{t}_0 \times \omega)\circ[3 \hat{t}_0\hat{t}_0 -2 \mathbb{I}]] ds \nonumber \\
&= \int_0^{\delta} s\begin{bmatrix} t_{0x} \\ t_{0y} \\ 0 \end{bmatrix}   \times \left[ \begin{bmatrix} -c v_{0x} \\ -c v_{0y} \\ 0 \end{bmatrix} + s\begin{bmatrix} t_{0x} \\ t_{0y} \\ 0 \end{bmatrix} \times \begin{bmatrix} 0 \\ 0 \\ \omega_0 \end{bmatrix} - log(2) c^2 \right. \nonumber \\
&\left. \:\: \times \left( \begin{bmatrix} v_{0x} \\ v_{0y} \\ 0 \end{bmatrix} + s\begin{bmatrix} t_{0x} \\ t_{0y} \\ 0 \end{bmatrix} \times \begin{bmatrix} 0 \\ 0 \\ \omega_0 \end{bmatrix} \right) \left( \circ \begin{bmatrix} t_{0x} \\ t_{0y} \\ 0 \end{bmatrix} \begin{bmatrix} t_{0x} \\ t_{0y} \\ 0 \end{bmatrix} -2\mathbb{I} \right) \right. \nonumber \\
& \left.  \:\: - \left( c^2 \begin{bmatrix} v_{0x} \\ v_{0y} \\ 0 \end{bmatrix} + s \begin{bmatrix} t_{0x} \\ t_{0y} \\ 0 \end{bmatrix} \times \begin{bmatrix} 0 \\ 0 \\ \omega_0 \end{bmatrix} \right) \right]  \nonumber \\
& \left. \:\: \left( \circ \begin{bmatrix}
3 \begin{bmatrix} t_{0x} \\ t_{0y} \\ 0 \end{bmatrix} \begin{bmatrix} t_{0x} \\ t_{0y} \\ 0 \end{bmatrix} -2\mathbb{I} \end{bmatrix} \right) \right]   ds \nonumber \\
&= \int_0^{\delta} \begin{bmatrix}
cst_{0y} - (2log(2) -2)st_{0y} \\ -sc t_{0x} + (2log(2) -2)st_{0x} \\ -s^2 t^2_{0x}-2s^2t_{0y}^2 - (2log(2) -2) s t_{0x}
\end{bmatrix}^T \begin{bmatrix}
v_{0x} \\
v_{0y} \\
\omega_0
\end{bmatrix} ds \nonumber \\
&= \begin{bmatrix}
a_{31} & a_{32} & a_{33}
\end{bmatrix} \begin{bmatrix}
v_{0x} \\
v_{0y} \\
\omega_0
\end{bmatrix} \label{Moment_LHS_final}
\end{align}
where in the last 2 equations just the third row of equations is written since the the first $2$ rows corresponding to the $x$ and $y$ components would be zero after taking the cross products. The terms $a_{ij}$ in these equations are 
\begin{align*}
& a_{31} = c \frac{\delta^2}{2} - (2log(2) -2)\frac{\delta^2}{2}t_y ,\\
& a_{32} = -\frac{\delta^2}{2}c t_x+(2log(2) -2)\frac{\delta^2}{2}t_x , \\
& a_{33} = -\frac{\delta^3}{3} t^2_x-2\frac{\delta^3}{3}t_y^2 - (2log(2) -2) \frac{\delta^2}{2}t_x
\end{align*}
Thus, combining equations \ref{Forces_LHS_final} and \ref{Moment_LHS_final}, we get 3 equations in $v_{0x}, v_{0y}, \omega_0$. These $3$ equations are the forces and moments acting on the swimmer head of length $\delta$, which are the left hand sides of equations \ref{Force1} and \ref{Moment1}. Substituting these in the left hand sides of equations \ref{Force1} and \ref{Moment1}, respectively, and combining the 2 equations, we get the following single equation
\begin{equation}
A(k, \hat{t}, \delta) \begin{bmatrix}
v_{0x} \\
v_{0y} \\
\omega_0
\end{bmatrix} = F_{tail} (u^*, t, k)
\end{equation}
where, $A(k, \hat{t}, \delta)$ is a $3\times3$ matrix such that $A(i,j) = a_{ij}$ and the total forces and moments acting on the tail is $F_{tail}$ which is the function of the shape velocity $u^* : (\delta,1] \in \mathbb{R} \mapsto \mathbb{R}^2 $ and the instantaneous shape of the body which is the space of all continuous curves in $\mathbb{R}^2$. Substituting $\xi = [v_{0x}, v_{0y}, \omega_0]^T$ as the resultant velocity of the swimmer head, we get the following kinematic equation
\begin{equation}\label{kinematic_form}
\xi = - A^{-1} (k, \hat{t}, \delta) F_{tail} (u^*, t, k)
\end{equation}
This is the expression for the head velocity of the slender, flexible microswimmer in the coordinate frame attached to the tip of the head. In the following section we shall have a look in detail the form of this equation and the topological structure of the  configuration space of this type of swimmers.

\section{Geometry of the configuration space}
\begin{figure}[htb!]
  \begin{center}
\includegraphics[scale=0.48]{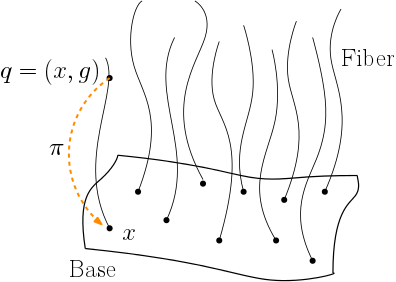}
\caption{Fiber Bundle}
\label{fiber_bundle}
  \end{center}
\end{figure}

The geometry of the configuration space requires attention for elegant and insightful solutions while studying the problem of locomotion using shape change. For such systems, the configuration space is usually written as the product of two manifolds. One is the base space or the shape space $M$ which describes the configuration of the internal shape variables of the mechanism, and the other is a Lie group $G$ which represents the macro-position of the locomoting body and is usually $SE(3)$ or one of its submanifolds. The total configuration space of the robot $Q$ then naturally appears as a product $M \times G$. Such systems follow the topology of a trivial principal fiber bundle, see \cite{kobayashi1963foundations}. Figure \ref{fiber_bundle} shows an explanatory figure of a fiber bundle. With such a separation of the configuration space, locomotion is readily seen as the means by which changes in shape affect the macro position. We refer to \cite{bloch1996nonholonomic}, \cite{kelly1995geometric} for a detailed explanation on the topology of locomoting systems. In particular, \cite{montgomery2006tour} discusses how the flexible locomotion systems such as the low Reynolds swimmers' configuration space admit this topological structure. We now show that the configuration space of a slender flexible swimmer is a trivial principal fiber bundle, defined as follows \cite{kelly1995geometric}.

\begin{definition}:
\textit{For $Q$ a configuration manifold and $G$ a Lie group, a trivial principal fiber bundle with base $M$ and structure group $G$ is a manifold $Q = M \times G$ with a free left action of $G$ on $Q$ given by left translation in the group variable: $\phi_h(x,g) = (x,hg)$ for $x \in M$ and $g \in G$}.
\end{definition}
We model the slender flexible swimmer of length $L$ as an embedding $\psi$ of interval $I = [0,L]$ into $\mathbb{R}^3$
\begin{equation}
\psi : I \mapsto \mathbb{R}^3
\end{equation}
The configuration space $Q$ for a swimming microorganism is the space $Q$ of embeddings of the cell membrane, represented by $I$, into $\mathbb{R}^3$. The structure group $G$ in our case is the Special Euclidean group $SE(3)$. The shape space for the organism is space of unlocated curves of finite length in $\mathbb{R}^3$, which is the quotient of $Q$ by the group $SE(3)$ of rigid motions acting on $Q$ by composition. Since all the points $q \in Q$ are represented by $(x, g)$ with $x \in M$ and $g \in G$,  $Q$ has global product structure of the form $M \times G$. Moreover, $SE(3)$ acts via the left action as a matrix multiplication, and has a single identity element, which is a $4 \times 4$ identity matrix. Hence left action of the group, defined by $\Phi_h : (x,g) \in Q \longrightarrow (x,hg)$ is free, for $x \in M$  and $\: h,g \in G$.
Thus, the configuration space of the slender flexible swimmer satisfies a trivial principal fiber bundle structure.


\subsection{Controllability analysis}
\begin{figure}[!htb]
\centering
\includegraphics[scale=.59]{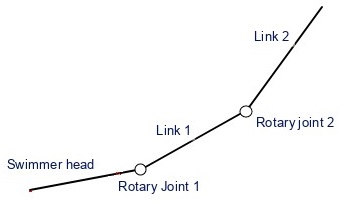}
\caption{Purcell's swimmer}
\label{Purcell_Jakarta}
\end{figure}

In this section we prove a result on the controllability of the flexible swimmer whose shape space $M$ is the space of all continuous curves in a plane. We refer to this swimmer as the $C^0$-planar flexible swimmer. The proof goes by construction through the result of strong controllability of the planar Purcell's swimmer already presented in the literature. As discussed in the introduction, the planar Purcell's swimmer is a the swimmer which has 3 straight, rigid and slender links with 2 rotary joints, see figure \ref{Purcell_Jakarta}. The angular velocity at these 2 joints is the control input to the system.
Clearly, the Purcell's swimmer is an example of a $C^0-$ flexible swimmer, whose shape variable takes values in the space of all the continuous curves having 3 interconnected line segments. This shape space can also be written as $SO(2) \times SO(2)$, which is of dimension 2. The system dynamics for the Purcell's swimmer can be written using the Cox theory itself in a driftless control affine form as follows \cite{hatton2013geometric}.
\begin{equation}\label{pure_kinematic2}
\begin{bmatrix}
\dot{x} \\
\xi
\end{bmatrix} = \begin{bmatrix}
I \\
-\mathbb{A}(x)
\end{bmatrix} u
\end{equation}
Here $\mathbb{A}(x)$ is the local connection form defined at each $x \in M$. We recall that for a shape space $M$, its tangent space at a point $x \in M$ is denoted by $T_xM$, and the shape velocity $\dot{x} = (\dot{\alpha}_1,\dot{\alpha}_2) \in T_xM$. Thus $\mathbb{A}(x) : TM \mapsto \mathfrak{g}$, is the local form of connection mapping the points in the tangent bundle $TM$ to the shape space $M$ to the Lie algebra $\mathfrak{g}$ of the structure group $G.$ $I$ is a $2 \times 2$ identity matrix and $u = [\dot{\alpha}_1, \dot{\alpha}_2]^T \in T_xM$ is the control input at each point $(\alpha_1, \alpha_2) \in SO(2) \times SO(2)$. The connection form and the other notions mentioned here have roots in geometric mechanics, see \cite{bloch2003nonholonomic}, \cite{holm2009geometric} for details.

Since our configuration space is naturally split into a shape and a structure group, we write a point in the configuration space as $q = (x,g) \in M \times G = Q$. We recall that for a time parametrized shape $x(t) \in M$, the horizontal lift $x^*(t) \in Q$ is a curve which projects to $x(t)$ under the projection map defining the principal fiber bundle and the components of its tangent vectors $\dot{x}^*(t) \in T_qQ$ satisfy the equation (\ref{pure_kinematic2}). We define following 2 controllability notions \cite{kelly1995geometric}.

\begin{itemize}
\item \textit{A locomotion system is said to be strongly controllable if, for any initial $q_0=(x_0,g_0)$ and final $q_f=(x_f,g_f)$, there exists a time $T > 0$ and a curve passing through $q_0$ satisfying $x^*(0) = q_0$ and $x^*(T) = q_f$.}

\item \textit{A locomotion system is said to be weakly controllable if, for any initial position $g_0 \in G$, and final position $g_f \in G$, and initial shape $x_0 \in M$, there exists a time $T > 0$ and a curve in the base space $x(t)$ satisfying $x(0) = x_0$ such that the horizontal lift of $x(t)$ passing through $(x_0, g_0)$  satisfies $x^*(0) = q_0$ and $x^*(T) = (x(T), g_f)$.}
\end{itemize}
The principal fiber bundle structure in controllability analysis gives rise to such strong and weak controllability notions which define finer ideas of controllability for locomotion systems. These notions are of practical relevance since many times just reaching the desired group component without strict requirement on shape of the system is sufficient. We define following vector spaces in terms of the local connection form $A(x)$, its curvature $DA(x)$, its Lie derivative $L_zDA(x)$ and their successive Lie brackets \cite{kelly1995geometric}
\begin{align*}
\mathfrak{h}_1\:&=\: span \{A(x)(X) : X \in T_x M\},\\
\mathfrak{h}_2\:&=\: span \{DA(x)(X,Y) : X, Y \in T_x M\}, \\
\mathfrak{h}_3\:&=\: span \{L_Z DA(x)(X,Y) - [A(x)(Z),DA(x)(X,Y)], \\ & \qquad [DA(x)(X,Y),DA(x)(W,Z)] : W, X, Y, Z \in T_xM\} \\
\vdots \\
\mathfrak{h}_k\:&=\: span \{L_X \xi - [A(x)(X),\xi],[\eta,\xi] :
X \in T_x M, \\
& \:\: \qquad\qquad \xi \in \mathfrak{h}_{k-1}, \eta \in \mathfrak{h}_2 \:\oplus\: \cdots \:\oplus\:\mathfrak{h}_{k-1}\}
\end{align*}
Then a system defined on a trivial principal bundle $Q$ is locally weakly controllable near $q \in Q$ if and only if the space of the Lie algebra $\mathfrak{g}$ of the structure group $G$ is spanned by the vector fields $\mathfrak{h}_1, \mathfrak{h}_2, \cdots$ as follows
\begin{equation}
\mathfrak{g}\:=\:\mathfrak{h}_1\:\oplus\:\mathfrak{h}_2\:\oplus \cdots 
\end{equation}
Whereas the system is locally strongly controllable if and only if 
\begin{equation}\label{weak_controllability}
\mathfrak{g}\:=\:\mathfrak{h}_2\:\oplus\:\mathfrak{h}_3\:\oplus \cdots 
\end{equation}
Using the analytical computations, the literature has already proven that for the planar Purcell's swimmer the rank of $\mathfrak{h}_2\:\oplus\:\mathfrak{h}_3$ is $3$ on the entire space $Q$, which is the dimension of the Lie algebra of its structure group $SE(2)$. Hence the swimmer satisfies the strong controllability conditions at all the points \cite{kadam2016geometric}, \cite{giraldi2013controllability}.

Now if we restrict the shape space of the flexible swimmer to all $C^0$ curves, then all the shapes that the rigid 3-link Purcell's swimmer can take are contained in the shape space $M$ of $C^0-$planar flexible swimmer. Moreover, the Purcell's swimmer mechanism when one of its outer link is considered as the head is similar to the $C^0$-planar flexible swimmer. Also, weak controllability notion concerns with this head's motion on $SE(2)$. The global strong controllability condition for the planar Purcell's swimmer means that the swimmer with 3 straight rigid links can be manoeuvred from any point $q_0$ to $q_1$ on its configuration space $Q$, where $q_0, q_1 \in Q = SO(2) \times SO(2) \times SE(2)$. Hence, the Lie algebra $\mathfrak{g}$ of the structure group $G$ of the $C^0-$planar flexible swimmer is spanned by the Lie algebra of the control vector fields, satisfying equation \ref{weak_controllability}. This proves that the $C^0-$planar flexible swimmer is weakly controllable.

We highlight that the strong controllability of the flexible swimmer is neither guaranteed nor ruled out by this approach of proof by construction since the shape space of the $C^0-$planar flexible swimmer by definition consists of all $C^0$ curves, of which the $C^0$ curves corresponding to non-straight link shapes are excluded in the proof. Nonetheless, the weak controllability result is of practical relevance since it means such a swimmer can be manoeuvred to achieve any arbitrary group displacement of its head on $SE(2)$.

\section{Simulation results}
\begin{figure}
\begin{minipage}[t]{0.49\linewidth}
    \includegraphics[width=0.86\linewidth]{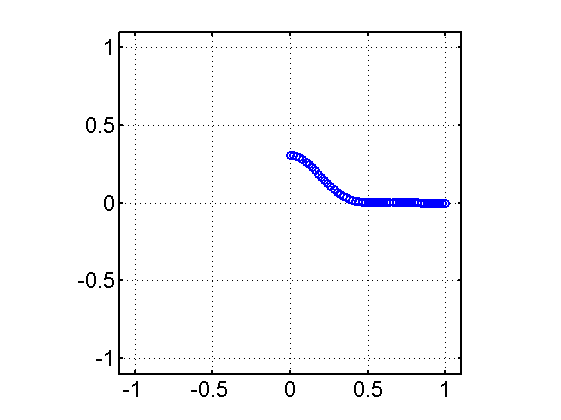}
    \subcaption{Time = 0 sec}
    \label{f1}
\end{minipage}
    \hfill
\begin{minipage}[t]{0.49\linewidth}
    \includegraphics[width=0.86\linewidth]{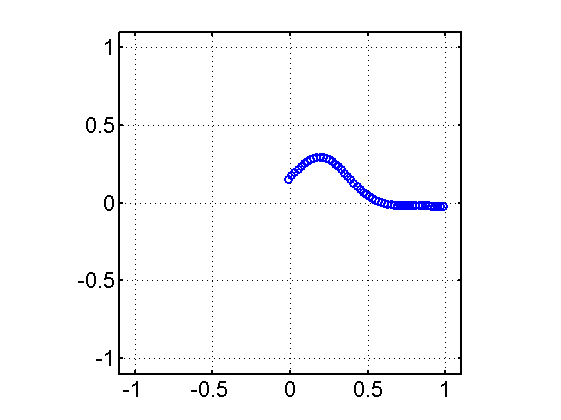}
    \subcaption{Time = 3 sec}
    \label{f2}
\end{minipage} 

\begin{minipage}[t]{0.49\linewidth}
    \includegraphics[width=0.86\linewidth]{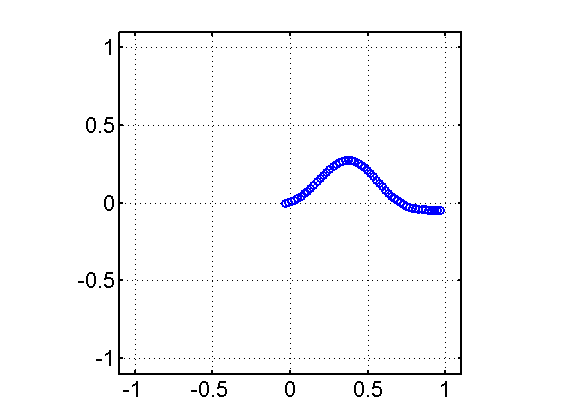}
    \subcaption{Time = 6 sec}
    \label{f3}
\end{minipage}
    \hfill
\begin{minipage}[t]{0.49\linewidth}
    \includegraphics[width=0.86\linewidth]{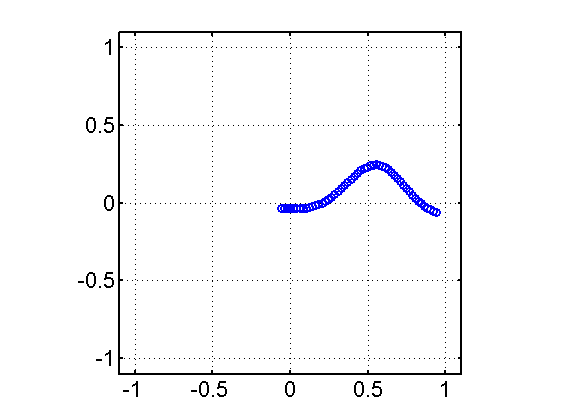}
    \subcaption{Time = 9 sec}
    \label{f4}
\end{minipage} 

\begin{minipage}[t]{0.49\linewidth}
    \includegraphics[width=0.86\linewidth]{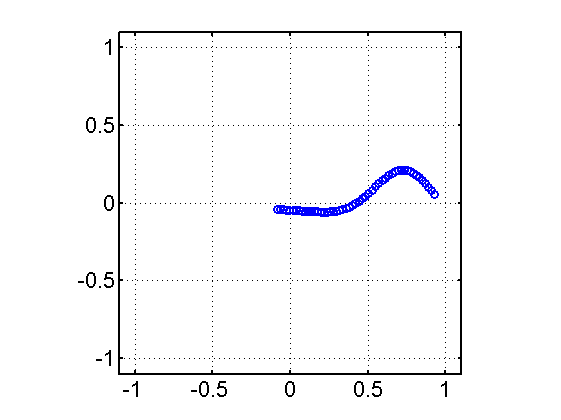}
    \subcaption{Time = 12 sec}
    \label{f5}
\end{minipage}
    \hfill
\begin{minipage}[t]{0.49\linewidth}
    \includegraphics[width=0.86\linewidth]{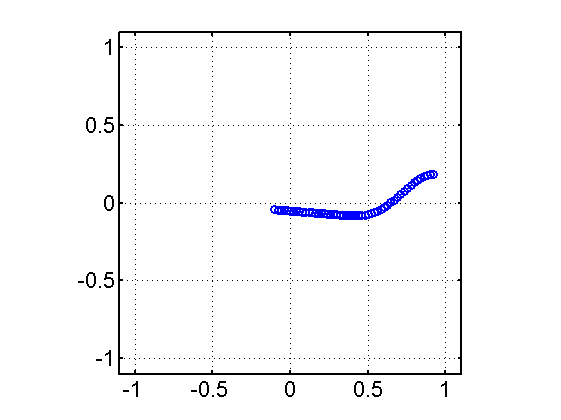}
    \subcaption{Time = 15 sec}
    \label{f6}
\end{minipage} 
\caption{Swimmer simulation against the moving bump function}
\end{figure}

\begin{figure}
\begin{minipage}[t]{0.45\linewidth}
\includegraphics[scale=0.24]{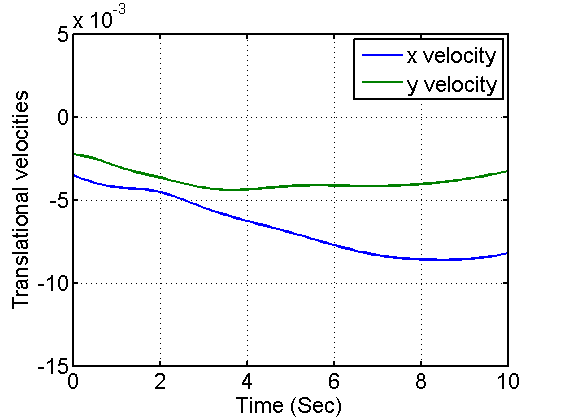}
\subcaption{Translational velocity}
\label{translational_velocity}
\end{minipage}
    \hfill
\begin{minipage}[t]{0.45\linewidth}
\includegraphics[scale=0.24]{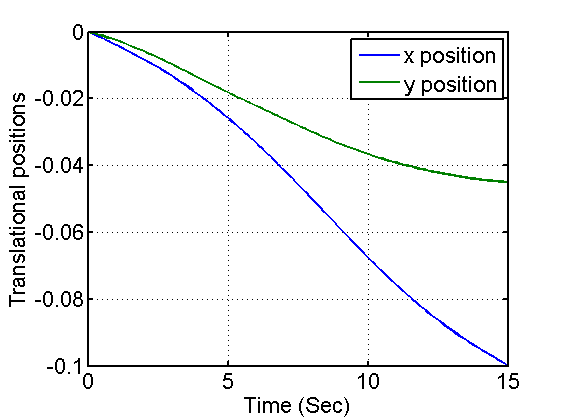}
\subcaption{Translational position}
\label{translational_position}
\end{minipage} 

\begin{minipage}[t]{0.45\linewidth}
\includegraphics[scale=0.24]{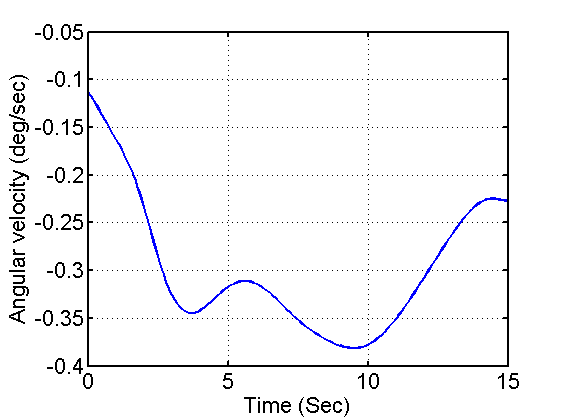}
\subcaption{Rotational velocity}
\label{rotational_velocity}
\end{minipage}
    \hfill
\begin{minipage}[t]{0.45\linewidth}
\includegraphics[scale=0.24]{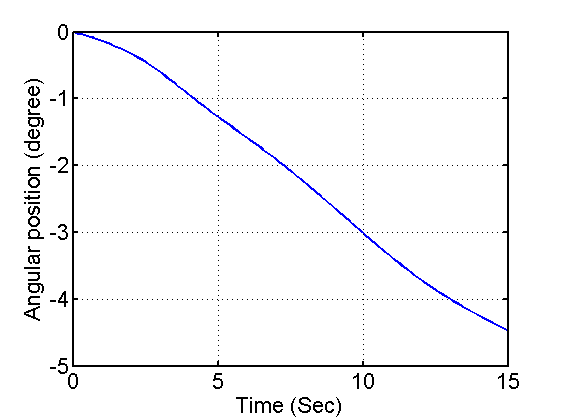}
\subcaption{Rotational position}
\label{rotational_position}
\end{minipage} 
\caption{Swimmer head motion response to the bump function}
\end{figure}

In this section we present the response of the head motion of the planar flexible swimmer subject to a bump function $\psi$, which defines displacement of the swimmer body in $y$ direction of the with respect to the head frame. The bump function travels from the swimmer head to tail over simulation period of $15$ seconds. 
\begin{equation}
{\displaystyle \Psi (x,t)={\begin{cases}c_1\exp \left({\frac {-c_2}{1-(x-c_3t)^{2}}}\right)&x\in (-1+c_3t,1+c_3t)\\0&{\mbox{ otherwise}}\end{cases}}}
\end{equation}
The values of the parameters of the bump function as $c_1 = 10^6, \:c_2 = 15, \:c_3 = \frac{1}{15}$. The resulting variation in the shape space is shown in figures \ref{f1} to \ref{f6}, where the head is at the left tip. This function defines the control input on the tangent bundle of the base space $TM$ of the swimmer, which includes the shape $\Psi$ and shape velocity $\dot{\Psi}$ of the swimmer. The swimmer head length $\delta$ is taken to be $0.05$ times the swimmer  length. The motion of the head is obtained using the kinematic equation \ref{kinematic_form}. The velocities and position of the head obtained are shown in figures \ref{translational_velocity} to \ref{rotational_position}, where quantities corresponding to the translational motion are dimensionless.

%
%
%

\section{CONCLUSIONS AND FUTURE WORK}
We have presented a kinematic model of a slender, flexible swimmer at low Reynolds number conditions. Cox theory is used to model the viscous forces as a function of the instantaneous shape and shape velocity of the swimmer. These forces along with the low inertia conditions are used to obtain the expression of the velocity of the swimmer head. We highlighted the peculiar topology of the principal fiber bundle that this swimmer admits. The notion of weak controllability is presented and it is shown that the planar, slender flexible micro-swimmer is weakly controllable for shape space consisting of the $C^0$ curves. An interesting avenue for the future work is to explore if the system is strong and weak controllable for $C^k$-smooth curves in the shape space. Also, given the weak controllability result in this paper, motion planning strategies for point to point reconfiguration is also a very relevant problem to explore and has applications in robotics and in understanding the microbial motion. 

\addtolength{\textheight}{-12cm}   



\bibliographystyle{IEEEtran}
\bibliography{bib_combined_IROS_IFACJSC_DMP}
\end{document}